\documentclass[a4paper]{jpconf}
\usepackage{graphicx}
\usepackage{epsfig}
\usepackage{float}
\usepackage{amsmath}
\begin{document}
\title{Asymptotic Grand Unification: The SO(10) case}

\author{Mohammed Omer Khojali$^{a,b,1}$, Alan S.~Cornell$^{a,2}$}
\address{$^{a}$Department of Physics, University of Johannesburg, PO Box 524, Auckland Park 2006, South Africa}
\address{$^{b}$Department of Physics, University of Khartoum, PO Box 321, Khartoum 11115, Sudan}

\author{Aldo Deandrea$^{c,d,3}$, Giacomo Cacciapaglia$^{c,d,4}$}
\address{$^{c}$Universit\'e de Lyon, Universit\'e Lyon 1, F-69622 Lyon, France}
\address{$^{d}$Institut de Physique des 2 Infinis (IP2I), UMR5822 CNRS/IN2P3, F-69622 Villeurbanne Cedex, France}
\author{Ammar Abdalgabar$^{e,5}$}
\address{$^{e}$University of Hafr Al Batin, college of Science, department of physics, Hafr Al Batin 39524, Kingdom of Saudi Arabia}
\author{Corentin Cot$^{f,6}$}
\address{$^{f}$Laboratoire de Physique des 2 Infinis (IJCLab), Université Paris-Saclay, Orsay, France}

\ead{ $^{1}$khogali11@gmail.com, $^{2}$acornell@uj.ac.za , $^{3}$deandrea@ipnl.in2p3.fr, $^{4}$g.cacciapaglia@ipnl.in2p3.fr, $^{5}$aabdalgabar@gmail.com, $^{6}$corentin.cot@ijclab.in2p3.fr }

\begin{abstract}
We explicitly test the asymptotic grand unification of a minimal 5-dimensional model with SO(10) gauge theory compactified on an $S^{1}/Z_{2}\times Z^{\prime}_{2}$ orbifold. We consider all matter fields as propagating in the bulk and show that the gauge couplings asymptotically run to a fixed point in the UV. However, the Yukawa couplings will typically hit a Landau pole right above the compactification scale in this class of SO(10) models.
\end{abstract}

\section{Introduction}

Theories of grand unification continue to play an important role in guiding the searches for extensions of the Standard Model (SM)~\cite{Bajc:2016efj,Cacciapaglia:2020qky}. The idea of grand unification theories (GUT) is to reduce all the gauge interactions to one single gauge group and all the fermionic multiplets into one or two different representations for each generation of matter~\cite{Csaki:2002ur,Hosotani:2015hoa,Khojali:2017rnl}. This single gauge group corresponds to a unification of the three forces described by the SM. Since our observations are mostly in agreement with a model based on the SM gauge group, we require that the unified gauge group has SM gauge group as a subgroup.
The SM group is rank 4, which means that the gauge group $G$ must be at least rank 4. In the same way as in the Higgs mechanism, we demand that the unified gauge group spontaneously breaks to the SM gauge group at some higher energy scale.
The $SO(10)$ group is a popular candidate for unification for many reasons; it contains both the Pati–Salam group,  $SU(5)\times U(1)$ (and hence also $SU(5)$) as subgroups and is therefore more ``unified" in a sense. It also embeds all SM fermions of a generation, plus the right-handed neutrino, into one single representation~ \cite{Dixit:1989ff,Mohapatra:1992dx} .
Since  $SO(10)$ is rank 5, which is one more than SM gauge group, there are several possibilities for symmetry breaking. On the one hand, this produces a rich variety of phenomenologies, but on the other hand, it introduces arbitrariness into the model in terms of the choice of scalar sector and potential. Furthermore, since the scalar and intermediate symmetry breaking steps affect the renormalization group (RG) running, the chosen breaking procedure can have an effect on the unification scale and hence the related phenomenology.

In this work, we shall study the non-supersymmetric extensions of the SM based on the gauge group $SO(10)$. In particular, we will study higher-dimensional non-supersymmetric orbifold models. We will consider a unification where the couplings unify asymptotically, as in these models with a compact extra dimension (which becomes relevant at scales higher than the electroweak (EW) scale) the gauge symmetry in the bulk is unified~\cite{Cacciapaglia:2020qky}. We study the asymptotic GUT based on an $SO(10)$ model in a flat $S^{1}/Z_{2}\times Z^{\prime}_{2}$ orbifold.

The structure of this paper is as follows: In section \ref{Sec:2} we outline the model setup, in section \ref{Sec:3} we explore the gauge running and asymptotic unification, and in section \ref{Sec:4} we present the running of the Yukawa couplings. In section \ref{Sec:5} we conclude.

\section{Model Setup}\label{Sec:2}

 We consider here a minimal $SO(10)$ grand unified model in five dimensions, where the extra dimension is compactified on an $S^{1}/ Z_{2} \times Z^{\prime}_{2}$ orbifold~\cite{Ohlsson:2018qpt,Meloni:2016rnt}.
 $SO(10)$ gauge symmetry is broken to a Pati-Salam model $SU(4)_{C}\times SU(2)_{L}\times SU(2)_{R}$  by an $Z_{2}\times Z_{2}^{\prime}$  orbifold twisting which generates two inequivalent fixed points.
One is the preserved $SO(10)$ symmetric fixed point (we call this the visible brane) while the other has only a Pati-Salam symmetry (we refer to as the PS hidden brane)~\cite{Bajc:2016eiw,Langacker:1980js}.

The breaking is performed with a scalar in the 16 or 126 representations of $SO(10)$, thus breaking to the $SU(5)$ is done by the ordinary Higgs mechanism on the brane (we will refer to this as brane breaking):
\begin{equation}
  \begin{split}
 16 \Rightarrow & 10+\overline{5}+1 \\
 126 \Rightarrow & 50+45+\overline{15}+10+\overline{5}+1 .
\end{split}
\end{equation}
Both the $16$ and $126$ representations contain a singlet under $SU(5)$, where we choose the adjoint scalar in the 16 representation. The minimal content at the $SO(10)$ scale to realise this symmetry breaking could be either 16 + $\overline{16}$ or 126 + 16. We could also use a 16 + 45 on the $SO(10)$ symmetric brane to break $SO(10)$ to $SU(5)$. The unbroken gauge group in the overlap of $SU(4)_C \times SU(2)_L \times SU(2)_R$ with $SU(5)$ is just the SM~\cite{Fukuyama:2002vv}.

\subsection{Boundary conditions}
All the quarks and leptons, including right-handed neutrinos, in each generation are unified to a single ${\bf 16}$ and $\overline{{\bf 16}}$ dimensional spinor representation field. One family of fermions with an addition of a right handed neutrino~\cite{Haba:2004xx,Fukuyama:2008pw} is: 
\begin{equation}
{\bf 16}\,= \,\left(t_{L}, \nu_{L}, b_{L}, \tau_{L}, B_{L}^{c}, \mathcal{T}_{L}^{c}, -T_{L}^{c}, -N_{L}^{c}, T_{R}, N_{R}, B_{R}, \mathcal{T}_{R}, b_{R}^{c}, \tau_{R}^{c}, -t_{R}^{c}, -\nu_{R}^{c} \right),
\end{equation}
and 
\begin{equation}
\overline{{\bf 16}}\,= \,\left(T_{L}, N_{L}, B_{L}, \mathcal{T}_{L}, b_{L}^{c}, \tau_{L}^{c}, -t_{L}^{c}, -\nu_{L}^{c}, t_{R}, \nu_{R}, b_{R}, \tau_{R}, B_{R}^{c}, \mathcal{T}_{R}^{c}, -T_{R}^{c}, -N_{R}^{c} \right).
\end{equation}
 We choose the $P_{0}$ and $P_{1}$ matrices to be:
\begin{equation}
\begin{split}
P_{0} = & \text{diag}(-1,-1,-1,+1,+1) \otimes \text{diag}(+1,+1),\\
P_{1} =& \text{diag}(+1,+1,+1,+1,+1) \otimes \text{diag}(+1,+1).
\end{split}
\end{equation}

{
\begin{table}[h!]
\centering
\begin{tabular}{|c| c|c|c|}
\hline 
Field &\ $(Z_{2}, Z_{2}^{\prime})$ &\ Zero Mode? &\ KK mass\\ [0.5ex]
\hline \hline
$l$ &$(+, +)$&  $\surd$& $2n/R$\\
$L$ &$(-, +)$& $-$& $(2n+1)/R$\\
\hline
$\tau$ &$(-, -)$&  $\surd$& $2n/R$\\
$\mathcal{T}$ &$(-, +)$ & $-$& $(2n+1)/R$\\
\hline
$q$ &$(+, +)$&  $\surd$& $2n/R$\\
$Q$ &$(-, +)$& $-$& $(2n+1)/R$\\
\hline
$t$ &$(-, -)$& $\surd$& $2n/R$\\
$T$ &$(-,+)$& $-$& $(2n+1)/R$\\
\hline
$b$ &$(-, -)$& $\surd$& $2n/R$\\
$B$ &$(-,+)$& $-$& $(2n+1)/R$\\
\hline
$N^{C}$ & $(-, -)$& $\surd$ & $2n/R$\\
\hline
$B_{\mu}$ &$(+, +)$& $\surd$& $2n/R$\\
$W^{a}_{\mu}$ &$(+, +)$& $\surd$& $2n/R$\\
$G_{\mu}^{i}$ &$(+, +)$& $\surd$& $2n/R$\\
\hline
$Y_{\alpha}, Y_{\alpha}^{\prime}$ & $(+, -)$& $-$ & $(2n+1)/R$\\
\hline
$A_{\alpha}, A_{\alpha}^{\prime}$ & $(+, -)$& $-$ & $(2n+1)/R$\\
\hline
$X_{\alpha}$ & $(+, -)$&  $-$ & $(2n+1)/R$\\
\hline
$W_{R}^{0}$ & $(+, -)$& $-$ & $(2n+1)/R$\\
$W_{R}^{+}$ & $(+, -)$&  $-$ & $(2n+1)/R$\\
$W_{R}^{-}$ & $(+, -)$&  $-$ & $(2n+1)/R$\\
\hline
$H_{10}$ & $(+, +)$&$\surd$& $2n/R$ \\
 & $(+, -)$& $-$& $(2n+1)/R$ \\
\hline
 \end{tabular}
\caption{\label{SM particle}\it Quantum numbers and parities of all the 5D fields.}
\end{table}
}

With the set of fields introduced in Table~\ref{SM particle}, we can write down the most general Lagrangian as:
 \begin{equation}
\begin{split}
\mathcal{L}_{\text{SO(10)}} = & -\frac{1}{4} F^{(a)}_{MN}\, F^{(a)MN} - \frac{1}{2\xi}\left(\partial_{\mu} A^{\mu} - \xi \partial_{5} A_{y} \right)^{2} + i \overline{{\bf 16}} {\not}D {\bf 16} \\ & + 
\sum_{a,b} Y_{10}^{ab}\,\overline{{\bf 16}}_{a}\,\Gamma^{i}\,{\bf 16}_{b} H^{i}_{10}  + |D_{M} H_{10}|^{2} - V(H_{10}),
\end{split}
 \end{equation}
 where ${\not}D \, =\, \Gamma^{M}\,D_{M}$, $\left(D_{M} = \partial_{M} - i\,g A_{M}^{ab}.\Sigma_{ab}\right)$, and $A_{M}^{ab}$ are real valued $N(N-1)/2$ vector gauge fields with $a,b = 1,2,....,10$, for the $SO(10)$ case. $\Sigma_{ab}$ are the antisymmetric representation matrices~\cite{King:2021gmj,Pica:2010xq}. Note that the inner product $\left( A_{\mu}^{ab}.\Sigma_{ab}\right)$ implies a sum over the group indices $ab$. Using $\Gamma_{i}'s$ we can construct the generators of $SO(10)$ group as:
 \begin{equation}
 \Sigma_{ij} = \frac{1}{4i}[\Gamma_{i}\,,\,\Gamma_{j}].
 \end{equation}

\section{Gauge running and asymptotic unification}\label{Sec:3}

\par First, in the SM the three SM gauge couplings run according to the following renormalization group equations (RGEs)~\cite{Abdalgabar:2017cjw}:

\begin{equation}
\left. 2\pi\frac{\text{d}\alpha_1}{\text{d}t} \right|_\text{SM} =  \frac{41}{10}\alpha_1^2 \,, \quad
\left. 2\pi\frac{\text{d}\alpha_2}{\text{d}t} \right|_\text{SM}  = -\frac{19}{6}\alpha_2^2 \,, \quad 
\left. 2\pi\frac{\text{d}\alpha_3}{\text{d}t} \right|_\text{SM}  = -7\alpha_3^2 \,, 
\end{equation}
where $\alpha_i = g_i^2/4\pi$. These equations will be used for the running of the Yukawa couplings between the EW scale and the compactification scale.
\\
Above the compactification scale, the couplings behave as a unified version and will all share the same gauge coefficient:
\begin{equation}
c_{SO(10)} = -\frac{11}{3} C_2(G) + \frac{1}{6} C_2(G) + \frac{4}{3}2 n_g (T_R(16)) + \frac{1}{3} n_H^{10} T_R(10),
\end{equation}
where each term refers to a different field (gauge 4-components + gauge fifth component + fermion + scalar), $n_g$ being the number of fermion generations in the bulk and $n_H^{10}$ being the number of $10$ Higgs in the theory. In $SO(10)$, we have $C_2(G) = N-2 = 8$, $T_R(16)$ = 2 and $T_R(10)$ = 1.
\par For one 10-Higgs and three bulk fermions families, we obtain
\begin{equation}
    c_{SO(10)} = -\frac{35}{3} \, .
\end{equation}
The RGEs then become: 
\begin{equation}
\begin{aligned}
2\pi\frac{\text{d}\alpha_1}{\text{d}t} & =  \frac{41}{10}\alpha_1^2 + \left(S(t) - 1\right)c_{SO(10)}\alpha_1^2 \,, \quad
\\
2\pi\frac{\text{d}\alpha_2}{\text{d}t} & = -\frac{19}{6}\alpha_2^2 + \left(S(t) - 1\right)c_{SO(10)}\alpha_2^2 \,, \quad 
\\
2\pi\frac{\text{d}\alpha_3}{\text{d}t} & = -7\alpha_3^2 + \left(S(t) - 1\right)c_{SO(10)}\alpha_3^2 \,.
\end{aligned}
\end{equation}

The generic structure of the running of the gauge couplings at one loop is given by:
 \begin{equation}\label{running}
2 \pi\ \frac{d\,\alpha_{i}}{d\,t} =b^{\rm SM}\,\alpha_{i}^{2} +\left(S(t)-1\right)\,b^{\rm aGUT}\,\alpha_{i}^{2},
\end{equation}
 where $t = \ln\left(\mu/M_{Z}\right)$ and contains the energy scale parameter $\mu$. We chose to use the Z mass as a reference scale, so that for $\mu = M_{Z}$ we have $t = 0$ and we can fix the initial conditions of the running~\cite{Cacciapaglia:2020qky}. The function $S(t)$, is defined as
  \begin{equation}
S(t) = \left\{ \begin{array}{l}
\mu\,R = M_Z\, R\ e^t \qquad  \mbox{for} \quad \mu > 1/R\,, \\
1\qquad \mbox{for} \quad M_Z < \mu < 1/R. 
\end{array} \right.
\end{equation}

The asymptotic behaviour of the running of the gauge couplings can be easily understood when rewriting Eq.~(\ref{running}) in terms of $\tilde{\alpha}$, \begin{equation} 
\frac{d \tilde{\alpha}}{d t} =\tilde{\alpha} + \frac{b^{aGUT}}{2\,\pi} (\tilde{\alpha})^2\,,
\end{equation}
where
\begin{equation}
\tilde{\alpha_{i}}(\mu)\,\sim\,\frac{\alpha_{i}(\mu)}{2}\,\mu\,R.
\end{equation}
As such we only retain the term proportional to $S(t)$ that grows with energy. The beta function vanishes at:
\begin{equation}
\tilde{\alpha}|_{\rm IV} = 0,\quad\quad\quad \tilde{\alpha}|_{\rm UV} = - \frac{2\,\pi}{b^{aGUT}}\,.
\end{equation}
Therefore, for three bulk generation we have
\begin{equation}
\tilde{\alpha}|_{\rm UV}\,=\,\frac{6\pi}{35}.
\end{equation}
The fixed point’s existence requires $n_{g}\,\le 5$. For $n_{g} = 6$, or more bulk generations, the asymptotic unification would fail.
\section{Running of the Yukawa couplings}\label{Sec:4}
First, in the SM the four Yukawa couplings run according to the following RGEs:
\begin{eqnarray}
\left. 2\pi\frac{\text{d}\alpha_t}{\text{d}t} \right|_\text{SM} &=& \left[ \frac{9}{2}\alpha_t + \frac{3}{2}\alpha_b + \alpha_{\tau} + \alpha_\nu  - \frac{17}{20}\alpha_1   - \frac{9}{4}\alpha_2 - 8\alpha_3 \right]\alpha_t \,, 
\\
\left. 2\pi\frac{\text{d}\alpha_b}{\text{d}t} \right|_\text{SM}  &=& \left[\frac{9}{2}\alpha_b + \frac{3}{2}\alpha_t + \alpha_{\tau} + \alpha_\nu
- \frac{1}{4}\alpha_1 - \frac{9}{4}\alpha_2-8\alpha_3\right]\alpha_b \,, \\
\left. 2\pi\frac{\text{d}\alpha_\tau}{\text{d}t} \right|_\text{SM}  &=& \left[\frac{5}{2}\alpha_{\tau} + 3\alpha_t + 3\alpha_b -\frac{1}{2} \alpha_\nu
- \frac{9}{4} (\alpha_1+\alpha_2)\right]\alpha_{\tau} \,, 
\\
\left. 2\pi\frac{\text{d}\alpha_\nu}{\text{d}t} \right|_\text{SM}  &=& \left[\frac{5}{2}\alpha_{\nu} + 3\alpha_t + 3\alpha_b -\frac{1}{2} \alpha_\tau
- \frac{9}{20} \alpha_1 - \frac{9}{4} \alpha_2\right]\alpha_{\nu} \,,
\end{eqnarray}
where $\alpha_f = y_f^2/4\pi$. These equations will be used for the running of the Yukawa couplings between the EW scale and the compactification scale.

\par Adding all the parts for the bulk top Yukawa beta function above the compactification scale, we obtain the following RGE with respect to:
\begin{equation}
\begin{aligned}
    \left.16 \pi^2 \frac{\text{d}Y}{\text{d}t}\right|_{\text{aGUT}} =&  y^3 \left(2\times8 +\frac{1}{2}\times10 +\frac{1}{2}\times10 + 2 \times 32 \right)
    \\
    & + y g_5^2 \left(6\times\left(-\frac{27}{8}\right) - 3\times\frac{9}{2} + 2\times \frac{27}{8} + \frac{1}{2} \times \frac{45}{8} + \frac{1}{2} \times \frac{45}{8} \right) 
    \\
    & + y g_5^2 \xi \left(2\times\left(-\frac{27}{8}\right) + \frac{45}{8} + \frac{45}{8} + \frac{9}{2} - 2\times\frac{9}{4} - 2\times \frac{9}{4} \right)  \; ,
\end{aligned}
\end{equation}
where, as explained before, we computed the coefficients of the gauge contribution with unified gauge coupling $g_5$. The sum of the contributions shows that the gauge-parameter $\xi$ vanishes, as expected by gauge-invariance. Then, the one loop RGEs give:
\begin{equation}
\begin{aligned}
\left.16 \pi^2 \frac{\text{d}Y}{\text{d}t}\right|_{\text{aGUT}} =&  90 y^3 - \frac{171}{8} y g_5^2.
\end{aligned}
\end{equation}
\begin{figure*}[t]
\centering
\includegraphics[width=0.47\textwidth]{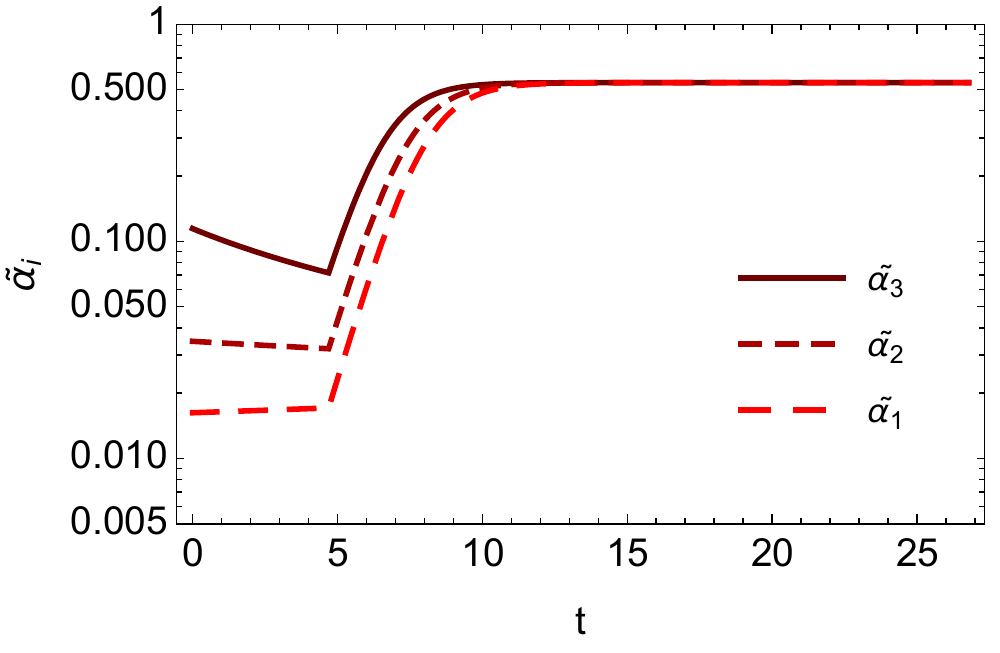}
\includegraphics[width=0.47\textwidth]{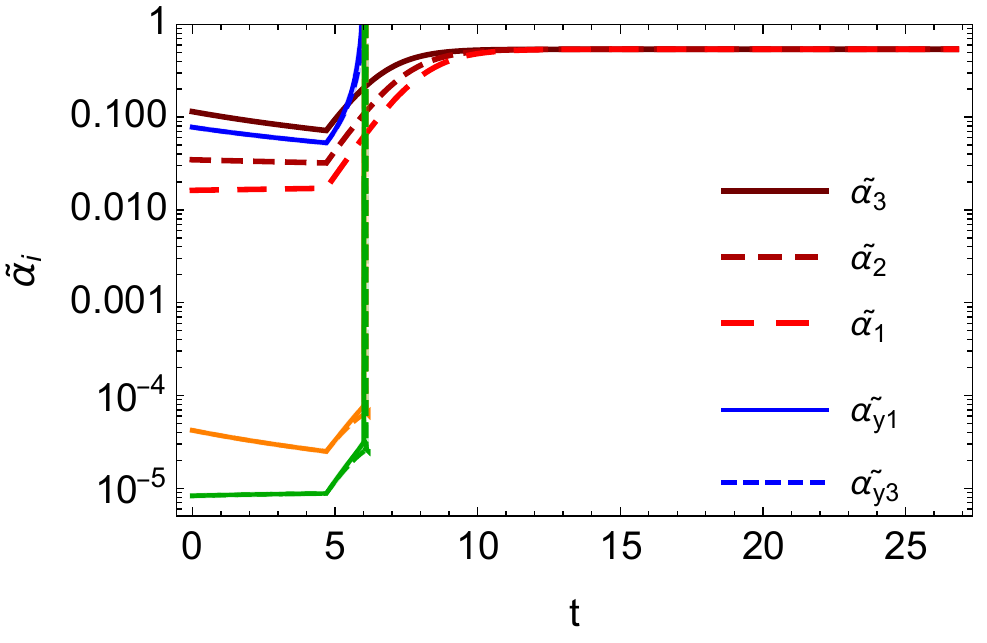}
\label{fig:eft-relic}
\caption{\small\it Running of the gauge couplings (left panel) and Yukawa coupling (right panel) using one-loop factors, with $R^{-1}=10$TeV. The range of $t$ corresponds to the $Z$ mass ($t=0$) and the reduced 5D Planck mass.  }
\label{fig1}
\end{figure*} 

Using the RGEs for 3 families of fermions we show in Fig.\ref{fig1} the one-loop evolution of
the three gauge couplings. We start the running at the $Z$ mass with the SM values $\{g^{0}_{1},g^{0}_{2},g^{0}_{3}\}=\{0.45,0.66,1.2\}$, while the
matching to the 5D running takes place at the scale $1/R$, indicated by the point where the running changes sharply~\cite{Cacciapaglia:2020qky}.
One can see that the couplings will never cross, therefore, they do get very close and tend to a unified value asymptotically at high energies. In fact, this value corresponds to the UV safe fixed point of the 5D theory. At $t\approx 9$ the couplings are effectively unified. In the case of the Yukawa couplings, they will typically hit a Landau pole before the GUT scale in this class of $SO(10)$ models.

\section{Conclusions}\label{Sec:5}
In this proceeding, we propose a 5-dimensional model that realizes asymptotic Grand Unification for the gauge couplings. The model is based on a bulk $SO(10)$ gauge theory compactified on an $S^{1}/ Z_{2} \times Z^{\prime}_{2}$ orbifold. We find that, in the minimal $SO(10)$ model, asymptotic unification is only possible for a number of bulk generations less than or equal to $5$. The gauge couplings asymptotically run to a unified fixed point in the UV. We also studied the running of the Yukawa couplings and found that the Yukawa couplings typically hit a Landau pole close to the compactification scale, hence reducing the validity of this model. 

\subsection*{Acknowledgments}
MOK was supported by the GES, and ASC is partially supported by the National Research Foundation South Africa.

\end{document}